\documentclass[10pt,conference]{IEEEtran}
\usepackage{epsfig,rotating,setspace,latexsym,amsmath,epsf,amssymb,bm,theorem}
\usepackage{cite}

\usepackage{amsfonts}

\title{Gaussian MIMO Broadcast Channels with Common and Confidential Messages
\thanks{This work was supported by NSF Grants CCF 04-47613, CCF 05-14846, CNS
07-16311 and CCF 07-29127.}}

\author{Ersen Ekrem \qquad Sennur Ulukus \\
\normalsize Department of Electrical and Computer Engineering\\
\normalsize University of Maryland, College Park, MD 20742 \\
\normalsize {\it ersen@umd.edu} \qquad {\it ulukus@umd.edu}}

\newcommand{\bbsigma}{\bm \Sigma}

\newcommand{\bbi}{{\mathbf{I}}}
\newcommand{\bzero}{{\mathbf{0}}}

\newcommand{\bbh}{{\mathbf{H}}}

\newcommand{\bbm}{{\mathbf{M}}}

\newcommand{\bbk}{{\mathbf{K}}}

\newcommand{\bbn}{{\mathbf{N}}}

\newcommand{\bbs}{{\mathbf{S}}}

\newcommand{\bbx}{{\mathbf{X}}}

\newcommand{\bby}{{\mathbf{Y}}}

\newtheorem{Theo}{Theorem}

\newtheorem{Lem}{Lemma}

\begin{document}

\IEEEoverridecommandlockouts

\maketitle

\begin{abstract}
We study the two-user Gaussian multiple-input multiple-output
(MIMO) broadcast channel with common and confidential messages. In
this channel, the transmitter sends a common message to both
users, and a confidential message to each user which is kept
perfectly secret from the other user. We obtain the entire
capacity region of this channel. We also explore the connections
between the capacity region we obtained for the Gaussian MIMO
broadcast channel with common and confidential messages and the
capacity region of its non-confidential counterpart, i.e., the
Gaussian MIMO broadcast channel with common and private messages,
which is not known completely.
\end{abstract}

\section{Introduction}
We study the two-user Gaussian multiple-input multiple-output
(MIMO) broadcast channel for the following scenario: The
transmitter sends a common message to both users, and a
confidential message to each user which needs to be kept perfectly
secret from the other user. We call the channel model arising from
this scenario the Gaussian MIMO broadcast channel with common and
confidential messages.

The Gaussian MIMO broadcast channel with common and confidential
messages subsumes many other channel models as special cases. The
first one is the Gaussian MIMO wiretap channel, where the
transmitter has only one confidential message for one (legitimate)
user, which is kept perfectly secret from the other user
(eavesdropper). The secrecy capacity of the Gaussian MIMO wiretap
channel is obtained in \cite{Hassibi,Wornell} for the general
case, in \cite{Ulukus} for the $2-2-1$ case. The second channel
model that the Gaussian MIMO broadcast channel with common and
confidential messages subsumes is the Gaussian MIMO wiretap
channel with common message~\cite{Liu_Common_Confidential}, in
which the transmitter sends a common message to both legitimate
user and the eavesdropper, and a confidential message to the
legitimate user that is kept perfectly secret from the
eavesdropper. The capacity region of the Gaussian MIMO wiretap
channel with common message is obtained
in~\cite{Liu_Common_Confidential}. The last channel model that the
Gaussian MIMO broadcast channel with common and confidential
messages encompasses is the the Gaussian MIMO broadcast channel
with confidential messages~\cite{Ruoheng_MIMO_BC}, where the
transmitter sends a confidential message to each user which is
kept perfectly secret from the other user. The capacity region of
the Gaussian MIMO broadcast channel with confidential messages is
established in~\cite{Ruoheng_MIMO_BC}.

Here, we obtain the capacity region of the Gaussian MIMO broadcast
channel with common and confidential messages. In particular, we
show that a variant of the secret dirty-paper coding (S-DPC)
scheme proposed in~\cite{Ruoheng_MIMO_BC} is capacity-achieving.
Since the S-DPC scheme proposed in~\cite{Ruoheng_MIMO_BC} is for
the transmission of only two confidential messages, it is modified
here to incorporate the transmission of a common message as well.
Similar to~\cite{Ruoheng_MIMO_BC}, we also notice an invariance
property of this achievable scheme with respect to the encoding
order used in the S-DPC scheme. In other words, two achievable
rate regions arising from two possible encoding orders used in the
S-DPC scheme are identical, and equal to the capacity region. We
provide the proof of this statement as well as the converse proof
for the capacity region of the Gaussian MIMO broadcast channel
with common and confidential messages by using channel enhancement
technique~\cite{Shamai_MIMO} and an extremal inequality
from~\cite{Liu_Compound}.

We also explore the connections between the Gaussian MIMO
broadcast channel with common and confidential messages and its
non-confidential counterpart, i.e., the (two-user) Gaussian MIMO
broadcast channel with common and private messages. In the
Gaussian MIMO broadcast channel with common and private messages,
the transmitter again sends a common message to both users, and a
private message to each user, for which there is no secrecy
constraint now, i.e., private message of each user does not need
to be kept secret from the other user. Thus, the Gaussian MIMO
broadcast channel with common and confidential messages we study
here can be viewed as a constrained version of the Gaussian MIMO
broadcast channel with common and private messages, where the
constraints come through forcing the private messages to be
confidential. We note that although there are partial results for
the capacity region of the Gaussian MIMO broadcast channel with
common and private messages~\cite{Hannan_Common,hannan_thesis}, it
is not known completely. However, here, we are able to obtain the
entire capacity region for a constrained version of the Gaussian
MIMO broadcast channel with common and private messages, i.e., for
the Gaussian MIMO broadcast channel with common and confidential
messages. We provide an intuitive explanation of this
at-first-sight surprising point  as well as the invariance
property of the achievable rate regions with respect to the
encoding orders that can be used in the S-DPC scheme by using a
result from~\cite{hannan_thesis} for the Gaussian MIMO broadcast
channel with common and private messages.


\section{Channel Model and Main Result}
We study the two-user Gaussian MIMO broadcast channel which is
defined by
\begin{align}
\bby_1&=\bbh_1 \bbx+\bbn_1 \label{general_gaussian_mimo_1}\\
\bby_2&=\bbh_2 \bbx+\bbn_2 \label{general_gaussian_mimo_2}
\end{align}
where the channel input $\bbx$ is a $t\times 1$ vector, $\bbh_j$
is the channel gain matrix of size $r_j\times t,$ the channel
output of the $j$th user $\bby_j$ is a $r_j\times 1$ vector, and
the Gaussian random vector $\bbn_j$ is of size $r_j\times 1$ with
a covariance matrix $\bbsigma_j$ which is assumed to be strictly
positive-definite, i.e., $\bbsigma_j\succ \bzero$. We consider a
covariance constraint on the channel input as follows
\begin{align}
E\left[\bbx\bbx^{\top}\right] \preceq \bbs
\label{covariance_constraint}
\end{align}
where $\bbs \succeq \bzero$.

We study the following scenario for the Gaussian MIMO broadcast
channel: There are three independent messages $(W_0,W_1,W_2)$ with
rates $(R_0,R_1,R_2)$, respectively, where $W_0$ is the common
message that needs to be delivered to both users, $W_1$ is the
confidential message of the first user which needs to be kept
perfectly secret from the second user, and similarly, $W_2$ is the
confidential message of the second user which needs to be kept
perfectly secret from the first user. The secrecy of the
confidential messages is measured by the normalized equivocation
rates~\cite{Wyner,Korner}, i.e, we require
\begin{align}
\frac{1}{n}I(W_1;W_0,W_2,\bby_{2}^n)\rightarrow 0~~{\rm and}~~
\frac{1}{n}I(W_2;W_0,W_1,\bby_{1}^n)\rightarrow 0
\label{equivocation_1}
\end{align}
as $n\rightarrow \infty$, where $n$ denotes the number of channel
uses. The closure of all achievable rate triples $(R_0,R_1,R_2)$
is defined to be the capacity region, and will be denoted by
$\mathcal{C}(\bbs)$. We next define the following shorthand
notations
\begin{align}
&R_{0j}(\bbk_1,\bbk_2)=\frac{1}{2} \log\frac{|\bbh_j \bbs
\bbh_j^{\top}+\bbsigma_j|}{|\bbh_j(\bbk_1+\bbk_2)\bbh_j^\top+\bbsigma_j|},\quad j=1,2 \label{common}\\
&R_1(\bbk_1,\bbk_2)=\frac{1}{2}\log\frac{|\bbh_1(\bbk_1+\bbk_2)\bbh_1^\top+\bbsigma_1|}{|\bbh_1
\bbk_2\bbh_1^\top+\bbsigma_1|}\nonumber\\
&\qquad \qquad \qquad
-\frac{1}{2}\log\frac{|\bbh_2(\bbk_1+\bbk_2)\bbh_2^\top+\bbsigma_2|}{|\bbh_2 \bbk_2\bbh_2^\top+\bbsigma_2|}\label{conf_1}\\
&R_2(\bbk_2)=
\frac{1}{2}\log\frac{|\bbh_2\bbk_2\bbh_2^\top+\bbsigma_2|}{|\bbsigma_2|}-\frac{1}{2}\log\frac{|\bbh_1\bbk_2\bbh_1^\top+\bbsigma_1|}{|\bbsigma_1|}
\label{conf_2}
\end{align}
using which, our main result can be stated as follows.
\begin{Theo}
\vspace{-0.25cm} \label{theorem_main_result} The capacity region
of the Gaussian MIMO broadcast channel with common and
confidential messages $\mathcal{C}(\bbs)$ is given by
\begin{align}
\mathcal{C}(\bbs)=\mathcal{R}_{12}^{\rm
S-DPC}(\bbs)=\mathcal{R}_{21}^{\rm S-DPC} (\bbs)
\end{align}
where $\mathcal{R}_{12}^{\rm S-DPC}(\bbs)$ is given by the union
of the rate triples $(R_0,R_1,R_2)$ satisfying
\begin{align}
R_0&\leq \min\{R_{01}(\bbk_1,\bbk_2),R_{02}(\bbk_1,\bbk_2)\}
\end{align}
\begin{align}
R_1&\leq R_1(\bbk_1,\bbk_2)\\
R_2&\leq R_2(\bbk_2)
\end{align}
for some positive semi-definite matrices $\bbk_1,\bbk_2$ such that
$\bbk_1+\bbk_2\preceq \bbs$, and $\mathcal{R}_{21}^{\rm
S-DPC}(\bbs)$ can be obtained from $\mathcal{R}_{12}^{\rm
S-DPC}(\bbs)$ by swapping the subscripts $1$ and 2.
\vspace{-0.25cm}
\end{Theo}
Theorem~\ref{theorem_main_result} states that the common message,
for which a covariance matrix $\bbs-\bbk_1-\bbk_2$ is allotted,
should be encoded by using a standard Gaussian codebook, and the
confidential messages, for which covariance matrices
$\bbk_1,\bbk_2$ are allotted, need to be encoded by using the
S-DPC scheme proposed in~\cite{Ruoheng_MIMO_BC}. The receivers
first decode the common message by treating the confidential
messages as noise, and then each receiver decodes the confidential
message intended to itself. Depending on the encoding order used
in S-DPC, one of the users gets a clean link for the transmission
of its confidential message, where there is no interference
originating from the other user's confidential message. Although
one might expect that two achievable regions arising from two
possible encoding orders that can be used in S-DPC could be
different, i.e., $\mathcal{R}_{12}^{\rm S-DPC}(\bbs)\neq
\mathcal{R}_{21}^{\rm S-DPC}(\bbs)$, and taking a convex closure
of these two regions would yield a larger achievable rate region,
Theorem~\ref{theorem_main_result} states that
$\mathcal{R}_{12}^{\rm S-DPC}(\bbs)= \mathcal{R}_{21}^{\rm
S-DPC}(\bbs)$, i.e., the achievable rate region is invariant with
respect to the encoding order used in S-DPC.

We conclude this section by defining a sub-class of Gaussian MIMO
broadcast channels called the aligned Gaussian MIMO broadcast
channel, which can be obtained from
(\ref{general_gaussian_mimo_1})-(\ref{general_gaussian_mimo_2}) by
setting $\bbh_1=\bbh_2=\bbi$, i.e.,
\begin{align}
\bby_1&=\bbx+\bbn_1\label{aligned_channel_1}\\
\bby_2&=\bbx+\bbn_2 \label{aligned_channel_2}
\end{align}
Here, we prove Theorem~\ref{theorem_main_result} for the aligned
channel. The proof for the general case in
(\ref{general_gaussian_mimo_1})-(\ref{general_gaussian_mimo_2})
can be carried out by using the capacity result for the aligned
case and the analysis in~\cite{MIMO_BC_Secrecy}.

\section{Proof of Theorem~\ref{theorem_main_result} for the Aligned Case}
Due to space limitations here, we omit the achievability proof for
Theorem~\ref{theorem_main_result}. We provide the converse proof.
Since the capacity region $\mathcal{C}(\bbs)$ is convex due to
time-sharing, it can be characterized by the solution of
\begin{align}
\max_{(R_0,R_1,R_2)\in\mathcal{C}(\bbs)} R_0+\mu_1 R_1+\mu_2 R_2
\end{align}
for $\mu_j\in[0,\infty),~j=1,2.$
To this end, we first characterize the boundary of
$\mathcal{R}_{12}^{\rm S-DPC}(\bbs)$ by studying the following
optimization problem
\begin{align}
\max_{(R_0,R_1,R_2)\in \mathcal{R}_{12}^{\rm S-DPC}(\bbs)}
R_0+\mu_1 R_1+\mu_2 R_2
\end{align}
which can be written as
\begin{align}
\max_{\substack{\bzero\preceq \bbk_j,~j=1,2
\\\bbk_1+\bbk_2\preceq \bbs}}~~&
\min\{R_{01}(\bbk_1,\bbk_2),R_{02}(\bbk_1,\bbk_2)\}\nonumber\\
& +\mu_1 R_1(\bbk_1,\bbk_2)+\mu_2 R_2(\bbk_2) \label{optimization}
\end{align}
Let $\bbk_1^*,\bbk_2^*$ be the maximizer of
$(\ref{optimization})$, which needs to satisfy the following KKT
conditions.
\begin{Lem}
\vspace{-0.25cm} \label{lemma_KKT_conditions} $\bbk_1^*,\bbk_2^*$
need to satisfy
\begin{align}
&\hspace{-0.25cm}(\mu_1+\mu_2)(\bbk_1^*+\bbk_2^*+\bbsigma_1)^{-1}+\bbm_1\nonumber\\
&\hspace{1.25cm}=(
\lambda+\mu_2)(\bbk_1^*+\bbk_2^*+\bbsigma_1)^{-1}\nonumber\\
&\hspace{1.25cm}\quad +(
\bar{\lambda}+\mu_1)(\bbk_1^*+\bbk_2^*+\bbsigma_2)^{-1}+\bbm_S \label{KKT_1}\\
&\hspace{-0.25cm}(\mu_1+\mu_2)(\bbk_2^*+\bbsigma_2)^{-1}+\bbm_2\nonumber\\
&\hspace{1.25cm}=(\mu_1+\mu_2)(\bbk_2^*+\bbsigma_1)^{-1}+\bbm_1\label{KKT_2}
\end{align}
for some positive semi-definite matrices $\bbm_1,\bbm_2,\bbm_S$
such that $
\bbk_1^*\bbm_1=\bbk_2^*\bbm_2=(\bbs-\bbk_1^*-\bbk_2^*)\bbm_S=\bzero\label{KKT_5}
$ and for some $\lambda=1-\bar{\lambda}$ such that it satisfies
$0\leq \lambda\leq 1$ and
\begin{align}
\lambda\left\{
\begin{array}{rcl}
=&0\qquad {\rm if}\quad R_{01}(\bbk_1^*,\bbk_2^*)>R_{02}(\bbk_1^*,\bbk_2^*)\\
=&1 \qquad {\rm if} \quad R_{01}(\bbk_1^*,\bbk_2^*)<R_{02}(\bbk_1^*,\bbk_2^*)\\
\neq & 0,1 \quad  {\rm if} \quad
R_{01}(\bbk_1^*,\bbk_2^*)=R_{02}(\bbk_1^*,\bbk_2^*)
\end{array}
\right.
\end{align}
\end{Lem}
Due to space limitations here, the proof of this lemma as well as
the proofs of the upcoming lemmas are omitted. We now use channel
enhancement~\cite{Shamai_MIMO} to define a new noise covariance
matrix $\tilde{\bbsigma}$ as follows
\begin{align}
(\mu_1+\mu_2)(\bbk_2^*+\tilde{\bbsigma})^{-1}&=(\mu_1+\mu_2)(\bbk_2^*+\bbsigma_2)^{-1}+\bbm_2
\label{def_enhanced_noise}
\end{align}
These new noise covariance matrix $\tilde{\bbsigma}$ has some
useful properties which are listed in the following lemma.
\begin{Lem}
\vspace{-0.25cm} \label{lemma_enhancement} We have the following
facts.
\begin{itemize}
\item $\tilde{\bbsigma}\preceq \bbsigma_1,$
$\tilde{\bbsigma}\preceq \bbsigma_2$.

\item $(\mu_1+\mu_2)(\bbk_1^*+\bbk_2^*+\tilde{\bbsigma})^{-1}=$

$\hspace{2.5cm}(\mu_1+\mu_2)(\bbk_1^*+\bbk_2^*+\bbsigma_1)^{-1}+\bbm_1$

\item
$(\bbk_2^*+\tilde{\bbsigma})^{-1}\tilde{\bbsigma}=(\bbk_2^*+\bbsigma_2)^{-1}\bbsigma_2$

\item
$(\bbk_1^*+\bbk_2^*+\tilde{\bbsigma})^{-1}(\bbk_2^*+\tilde{\bbsigma})=$

$\hspace{3cm}(\bbk_1^*+\bbk_2^*+\bbsigma_1)^{-1}(\bbk_2^*+\bbsigma_1)$
\end{itemize}
\end{Lem}
We now construct an enhanced channel using the new covariance
matrix $\tilde{\bbsigma}$ as follows
\begin{align}
\tilde{\bby}_1=\tilde{\bby}_2=\tilde{\bby}&=\bbx+\tilde{\bbn} \label{enhanced_channel_1} \\
\bby_1&=\bbx+\bbn_1 \label{enhanced_channel_3}\\
\bby_2&=\bbx+\bbn_2 \label{enhanced_channel_4}
\end{align}
where $\tilde{\bbn}$ is a Gaussian random vector with the
covariance matrix $\tilde{\bbsigma}$. In the channel given by
(\ref{enhanced_channel_1})-(\ref{enhanced_channel_4}), the
enhanced first and second users have the same observation
$\tilde{\bby}$. For the enhanced channel in
(\ref{enhanced_channel_1})-(\ref{enhanced_channel_4}), we consider
the scenario that a common message $W_0$ with rate $R_0$ is
directed to the first and second users, i.e., the users with
observations $\bby_1$ and $\bby_2$, respectively, $W_1$ (resp.
$W_2$) with rate $R_1$ (resp. $R_2$) is the confidential message
of the enhanced first (resp. second) user which is kept perfectly
hidden from the second (resp. first) user. We denote the capacity
region by $\tilde{C}(\bbs)$. Since in the enhanced channel, the
receivers to which only the common message is sent are identical
to the receivers in the original channel in
(\ref{aligned_channel_1})-(\ref{aligned_channel_2}), and the
receivers to which confidential messages are sent have better
observations with respect to the receivers in the original channel
in (\ref{aligned_channel_1})-(\ref{aligned_channel_2}), we have
$\mathcal{C}(\bbs)\subseteq \tilde{C}(\bbs)$. We next introduce an
outer bound for $\tilde{C}(\bbs)$.
\begin{Lem}
\vspace{-0.4cm} \label{lemma_outer_bound} The capacity region of
the enhanced channel in
(\ref{enhanced_channel_1})-(\ref{enhanced_channel_4})
$\tilde{C}(\bbs)$ is contained in the union of the rate triples
$(R_0,R_1,R_2)$ satisfying
\begin{align}
R_0&\leq \min\{I(U;\bby_1),I(U;\bby_2)\} \\
R_1&\leq I(\bbx;\tilde{\bby}|U)-I(\bbx;\bby_2|U)\\
R_2&\leq I(\bbx;\tilde{\bby}|U)-I(\bbx;\bby_1|U)
\end{align}
for some $(U,\bbx)$ such that $U\rightarrow \bbx \rightarrow
\tilde{\bby}\rightarrow (\bby_1,\bby_2)$ and
$E\left[\bbx\bbx^\top\right]\preceq \bbs$. \vspace{-0.1cm}
\end{Lem}
We also introduce the following extremal inequality
from~\cite{Liu_Compound}:
\begin{Lem}[\!\!\cite{Liu_Compound}, Corollary~4]
\vspace{-0.4cm} \label{lemma_extremal_liu} $(U,\bbx)$ is an
arbitrary random vector, where $E\left[\bbx\bbx^{\top}\right]
\preceq \bbs$ and $\bbs\succ \bzero$. Let
$\tilde{\bbn},\bbn_1,\bbn_2$ be Gaussian with covariance matrices
$\tilde{\bbsigma},\bbsigma_1,\bbsigma_2$, respectively. They are
independent of $(U,\bbx)$. Moreover,
$\tilde{\bbsigma},\bbsigma_1,\bbsigma_2$ satisfy
$\tilde{\bbsigma}\preceq \bbsigma_j,~j=1,2$. Assume that there
exists a covariance matrix $\bbk^*$ such that $\bbk^*\preceq \bbs$
and
\begin{align}
\beta(\bbk^*+\tilde{\bbsigma})^{-1}=\sum_{j=1}^2\gamma_j
(\bbk^*+\bbsigma_j)^{-1}+\bbm_S
\end{align}
where $\beta\geq 0, \gamma_j\geq 0,~j=1,2$ and $\bbm_S$ is
positive semi-definite matrix such that
$(\bbs-\bbk^*)\bbm_S=\bzero$. Then, for any $(U,\bbx)$, we have
\begin{align}
&\beta h(\bbx+\tilde{\bbn}|U)-\sum_{j=1}^2\gamma_j
h(\bbx+\bbn_j|U) \leq \nonumber\\
& \hspace{-0.15cm}\frac{\beta}{2} \log |(2\pi
e)(\bbk^*+\tilde{\bbsigma})| -\sum_{j=1}^2\frac{\gamma_j}{2} \log
|(2\pi e)(\bbk^*+\bbsigma_j)|
\end{align}
\vspace{-0.4cm}
\end{Lem}
We now use this lemma. For that purpose, we note that using the
second statement of Lemma~\ref{lemma_enhancement} in (\ref{KKT_1})
yields
\begin{align}
&\hspace{-0.25cm}(\mu_1+\mu_2)(\bbk_1^*+\bbk_2^*+\tilde{\bbsigma})^{-1}=(\lambda+\mu_2)(\bbk_1^*+\bbk_2^*+\bbsigma_1)^{-1}
\nonumber\\
&\qquad \qquad \qquad  +(
\bar{\lambda}+\mu_1)(\bbk_1^*+\bbk_2^*+\bbsigma_2)^{-1}+\bbm_S
\end{align}
using which in conjunction with Lemma~\ref{lemma_extremal_liu}, we
get
\begin{align}
&(\mu_1+\mu_2)h(\tilde{\bby}|U)-(\lambda+\mu_2)h(\bby_1|U)-(\bar{\lambda}+\mu_1)h(\bby_2|U)
\nonumber \\
&\leq \frac{\mu_1+\mu_2}{2} \log |(2\pi
e)(\bbk_1^*+\bbk_2^*+\tilde{\bbsigma})|\nonumber\\
&\quad -\frac{\lambda+\mu_2}{2}
\log|(2\pi e)(\bbk_1^*+\bbk_2^*+\bbsigma_1)| \nonumber \\
&\quad -\frac{\bar{\lambda}+\mu_1}{2} \log |(2\pi
e)(\bbk_1^*+\bbk_2^*+\bbsigma_2)| \label{extremal_result}
\end{align}
which will be used subsequently. We are now ready to complete the
converse proof as follows
\begin{align}
&\max_{(R_0,R_1,R_2)\in\mathcal{C}(\bbs)} R_0+\mu_1
R_1+\mu_2 R_2\\
&\leq \max_{(R_0,R_1,R_2)\in\tilde{\mathcal{C}}(\bbs)}
R_0+\mu_1 R_1+\mu_2 R_2 \label{outer_bound} \\
&\leq \max \hspace{0.1cm}  \min\{I(U;\bby_1),I(U;\bby_2)\}+
(\mu_1+\mu_2) I(\bbx;\tilde{\bby}|U)
\nonumber\\
&\qquad \qquad  -\mu_1 I(\bbx;\bby_2|U) -\mu_2 I(\bbx;\bby_1|U) \label{outer_bound_1}
\end{align}
\begin{align}
&\leq \max  ~\lambda I(U;\bby_1)+\bar{\lambda }I(U;\bby_2)+
(\mu_1+\mu_2)
I(\bbx;\tilde{\bby}|U) \nonumber\\
&\qquad \qquad  -\mu_1I(\bbx;\bby_2|U) -\mu_2 I(\bbx;\bby_1|U)
\label{non_negativity} \\
&= \max~ \lambda h(\bby_1)+\bar{\lambda
}h(\bby_2)+(\mu_1+\mu_2)h(\tilde{\bby}|U)\nonumber\\
&\qquad \quad -(\lambda+\mu_2)h(\bby_1|U)
-(\bar{\lambda}+\mu_1)h(\bby_2|U) \nonumber\\
&\qquad \quad -\frac{\mu_1}{2}
\log\frac{|\tilde{\bbsigma}|}{|\bbsigma_2|}-\frac{\mu_2}{2}
\log\frac{|\tilde{\bbsigma}|}{|\bbsigma_1|} \\
&\leq \frac{\lambda}{2} \log |(2\pi e)(\bbs+\bbsigma_1)|
+\frac{\bar{\lambda }}{2}  \log|(2\pi
e)(\bbs+\bbsigma_1)|\nonumber\\
&\quad + \max \left[
(\mu_1+\mu_2)h(\tilde{\bby}|U)-(\lambda+\mu_2)h(\bby_1|U)\right.\nonumber\\
& \quad -(\bar{\lambda}+\mu_1)h(\bby_2|U)\Big] -\frac{\mu_1}{2}
\log\frac{|\tilde{\bbsigma}|}{|\bbsigma_2|}-\frac{\mu_2}{2}
\log\frac{|\tilde{\bbsigma}|}{|\bbsigma_1|} \label{max_entropy_theorem} \\
&\leq \frac{\lambda}{2} \log |(2\pi e)(\bbs+\bbsigma_1)|
+\frac{\bar{\lambda }}{2}  \log|(2\pi
e)(\bbs+\bbsigma_1)|\nonumber\\
&\quad + \frac{(\mu_1+\mu_2)}{2} \log |(2\pi
e)(\bbk_1^*+\bbk_2^*+\tilde{\bbsigma})|\nonumber\\
&\quad -\frac{(\lambda+\mu_2)}{2}
\log |(2\pi e)(\bbk_1^*+\bbk_2^*+\bbsigma_1)| \nonumber \\
&\quad -\frac{\bar{\lambda}+\mu_1}{2}\log|(2\pi
e)(\bbk_1^*+\bbk_2^*+\bbsigma_2)| \nonumber\\
&\quad -\frac{\mu_1}{2}
\log\frac{|\tilde{\bbsigma}|}{|\bbsigma_2|}-\frac{\mu_2}{2}
\log\frac{|\tilde{\bbsigma}|}{|\bbsigma_1|}
\label{extremal_result_implies}\\
&=
\min\{R_{01}(\bbk_1^*,\bbk_2^*),R_{02}(\bbk_1^*,\bbk_2^*)\}\nonumber\\
&\quad  + \frac{\mu_1}{2} \log
\frac{|(\bbk_1^*+\bbk_2^*+\tilde{\bbsigma})\bbsigma_2|}{|(\bbk_1^*+\bbk_2^*+\bbsigma_2)\tilde{\bbsigma}|}
\nonumber\\
&\quad + \frac{\mu_2}{2} \log
\frac{|(\bbk_1^*+\bbk_2^*+\tilde{\bbsigma})\bbsigma_1|}{|(\bbk_1^*+\bbk_2^*+\bbsigma_1)\tilde{\bbsigma}|}\\
&= \min\{R_{01}(\bbk_1^*,\bbk_2^*),R_{02}(\bbk_1^*,\bbk_2^*)\}
+\mu_1 R_1(\bbk_1^*,\bbk_2^*)\nonumber\\
&\quad+\mu_2 R_2(\bbk_2^*)\label{lemma_enhancement_implies_5}
\end{align}
where (\ref{outer_bound}) comes from the fact that
$\mathcal{C}(\bbs)\subseteq \tilde{\mathcal{C}}(\bbs)$,
(\ref{outer_bound_1}) is due to Lemma~\ref{lemma_outer_bound},
(\ref{non_negativity}) results from the fact that $0\leq
\lambda=1-\bar{\lambda} \leq 1$, (\ref{max_entropy_theorem}) is
due to the maximum entropy theorem,
(\ref{extremal_result_implies}) comes from
(\ref{extremal_result}), and (\ref{lemma_enhancement_implies_5})
will be shown next. We first note the following
\begin{align}
R_1(\bbk_1^*,\bbk_2^*)&=\frac{1}{2} \log
\frac{|(\bbk_1^*+\bbk_2^*+\bbsigma_1)(\bbk_1^*+\bbk_2^*+\bbsigma_2)^{-1}|}{|(\bbk_2^*+\bbsigma_1)(\bbk_2^*+\bbsigma_2)^{-1}|}
\\
&=\frac{1}{2} \log
\frac{|(\bbk_1^*+\bbk_2^*+\tilde{\bbsigma})(\bbk_1^*+\bbk_2^*+\bbsigma_2)^{-1}|}{|(\bbk_2^*+\tilde{\bbsigma})(\bbk_2^*+\bbsigma_2)^{-1}|}\label{lemma_enhancement_implies_1}\\
&= \frac{1}{2}
 \log
\frac{|(\bbk_1^*+\bbk_2^*+\tilde{\bbsigma})\bbsigma_2|}{|(\bbk_1^*+\bbk_2^*+\bbsigma_2)\tilde{\bbsigma}|}
\label{lemma_enhancement_implies_2}
\end{align}
where (\ref{lemma_enhancement_implies_1}) is due to the fourth
statement of Lemma~\ref{lemma_enhancement} and
(\ref{lemma_enhancement_implies_2}) comes from the third statement
of Lemma~\ref{lemma_enhancement}. We next note the following
identity
\begin{align}
R_2(\bbk_2^*)&=\frac{1}{2}\log
\frac{|(\bbk_2^*+\bbsigma_2)(\bbk_2^*+\bbsigma_1)^{-1}|}{|\bbsigma_2
\bbsigma_1^{-1}|}\\
&=\frac{1}{2}\log
\frac{|(\bbk_2^*+\tilde{\bbsigma})(\bbk_2^*+\bbsigma_1)^{-1}|}{|\tilde{\bbsigma}
\bbsigma_1^{-1}|} \label{lemma_enhancement_implies_3}
\end{align}
\begin{align}
&= \frac{1}{2}\log
\frac{|(\bbk_1^*+\bbk_2^*+\tilde{\bbsigma})\bbsigma_1|}{|(\bbk_1^*+\bbk_2^*+\bbsigma_1)\tilde{\bbsigma}|}
\label{lemma_enhancement_implies_4}
\end{align}
where (\ref{lemma_enhancement_implies_3}) is due to the third
statement of Lemma~\ref{lemma_enhancement}, and
(\ref{lemma_enhancement}) comes from the fourth statement of
Lemma~\ref{lemma_enhancement}. Identities in
(\ref{lemma_enhancement_implies_2}) and
(\ref{lemma_enhancement_implies_4}) give
(\ref{lemma_enhancement_implies_5}). Thus, in the view of
(\ref{lemma_enhancement_implies_5}), we have shown that
$\mathcal{C}(\bbs)=\mathcal{R}_{12}^{\rm S-DPC}(\bbs)$. Similarly,
one can also show $\mathcal{C}(\bbs)=\mathcal{R}_{21}^{\rm
S-DPC}(\bbs)$; completing the proof of
Theorem~\ref{theorem_main_result}.

\section{Connections to the Gaussian MIMO Broadcast Channel with Common and Private Messages}
Here, we provide an intuitive explanation for the two facts that
Theorem~\ref{theorem_main_result} reveals: i) The achievable rate
region does not depend on the encoding order used in S-DPC, i.e.,
$\mathcal{R}_{12}^{\rm S-DPC}(\bbs)=\mathcal{R}_{21}^{\rm
S-DPC}(\bbs)$, ii) The capacity region of the Gaussian MIMO
broadcast channel with common and confidential messages can be
completely characterized, although the capacity region of the its
non-confidential counterpart, i.e., the Gaussian MIMO broadcast
channel with common and private messages, is not known completely.

In the Gaussian MIMO broadcast channel with common and private
messages, there are again three messages $W_0,W_1,W_2$ with rates
$R_0,R_1,R_2$, respectively, such that $W_0$ is again sent to both
users, $W_1$ (resp. $W_2$) is again directed to only the first
(resp. second) user, however, there are no secrecy constraints on
$W_1,W_2$. The capacity region of the Gaussian MIMO broadcast
channel with common and private messages will be denoted by
$\mathcal{C}^{\rm NS}(\bbs)$. The achievable rate region for the
Gaussian MIMO broadcast channel with common and private messages
that can be obtained by using DPC will be denoted by
$\mathcal{R}_{12}^{\rm NS-DPC}(\bbs),\mathcal{R}_{21}^{\rm
NS-DPC}(\bbs)$ (depending on the encoding order), where
$\mathcal{R}_{12}^{\rm NS-DPC}(\bbs)$ is given by the rate triples
$(R_0,R_1,R_2)$ satisfying
\begin{align}
R_0&\leq \min\{R_{01}^{\rm NS}(\bbk_1,\bbk_2),R_{02}^{\rm
NS}(\bbk_1,\bbk_2)\} \\
R_1&\leq R_{1}^{\rm NS}(\bbk_1,\bbk_2) \\
R_2&\leq R_{2}^{\rm NS}(\bbk_2)
\end{align}
for some positive semi-definite matrices $\bbk_1,\bbk_2$ such that
$\bbk_1+\bbk_2\preceq \bbs$, and $ \{R_{0j}^{\rm
NS}(\bbk_1,\bbk_2)\}_{j=1}^2,\break R_{1}^{\rm
NS}(\bbk_1,\bbk_2),R_{2}^{\rm NS}(\bbk_2)$ are defined as
\begin{align}
R_{0j}^{\rm NS}(\bbk_1,\bbk_2)&=\frac{1}{2}
\log\frac{|\bbs+\bbsigma_j|}{|\bbk_1+\bbk_2+\bbsigma_j|},\quad
j=1,2 \\
R_{1}^{\rm NS}(\bbk_1,\bbk_2)&=\frac{1}{2} \log
\frac{|\bbk_1+\bbk_2+\bbsigma_1|}{|\bbk_2+\bbsigma_1|} \\
R_{2}^{\rm NS}(\bbk_2)&=\frac{1}{2} \log
\frac{|\bbk_2+\bbsigma_2|}{|\bbsigma_2|}
\end{align}
Moreover, $\mathcal{R}_{21}^{\rm NS-DPC}(\bbs)$ can be obtained
from $\mathcal{R}_{12}^{\rm NS-DPC}(\bbs)$ by swapping the
subscripts 2 and 1. This achievable rate region was proposed
in~\cite{Goldsmith_common}.

We now state a result of~\cite{hannan_thesis} on the capacity
region of the Gaussian MIMO broadcast channel with common and
private messages, which is that for a given common message rate
$R_0$, the private messages sum rate capacity, i.e., $R_1+R_2$, is
achieved by both $\mathcal{R}_{12}^{\rm NS}(\bbs)$ and
$\mathcal{R}_{21}^{\rm NS}(\bbs)$. This result can also be stated
as follows
\begin{align}
&\max_{(R_0,R_1,R_2)\in\mathcal{C}^{\rm NS}(\bbs)} \mu_0^\prime
R_0+ \mu_1^{\prime} R_1 +\mu_2^{\prime} R_2 \nonumber\\
&=\max_{(R_0,R_1,R_2)\in \mathcal{R}_{12}^{\rm NS-DPC}(\bbs)}
\mu_0^\prime R_0+ \mu_1^{\prime} R_1 +\mu_2^{\prime} R_2 \label{hannan_did_it_1}\\
&= \max_{(R_0,R_1,R_2)\in \mathcal{R}_{21}^{\rm NS-DPC}(\bbs)}
\mu_0^\prime R_0+ \mu_1^{\prime} R_1 +\mu_2^{\prime} R_2
\label{hannan_did_it_2}
\end{align}
for $\mu_1^\prime=\mu_2^\prime=\mu^\prime$.  This result is
crucial to understand the aforementioned two facts suggested by
Theorem~\ref{theorem_main_result}, which will be explained next
using (\ref{hannan_did_it_1})-(\ref{hannan_did_it_2}).

In the proof of Theorem~\ref{theorem_main_result}, first, we
characterize the boundary of $\mathcal{R}_{12}^{\rm S-DPC}(\bbs)$
by finding the properties of the covariance matrices that achieve
the boundary of $\mathcal{R}_{12}^{\rm S-DPC}(\bbs)$, see
Lemma~\ref{lemma_KKT_conditions}. According to
Lemma~\ref{lemma_KKT_conditions}, the boundary of
$\mathcal{R}_{12}^{\rm S-DPC}(\bbs)$ can be achieved by using the
covariance matrices $\bbk_1^*,\bbk_2^*$ satisfying
(\ref{KKT_1})-(\ref{KKT_2}). On the other hand, using these
covariance matrices, we can also achieve the boundary points of
$\mathcal{R}_{12}^{\rm NS-DPC}(\bbs)$, which are actually on the
boundary of the capacity region $\mathcal{C}^{\rm NS}(\bbs)$ as
well, and are the private message sum rate capacity points for a
given common message rate. To see this point, we define
$\mu^\prime=\mu_1+\mu_2,\mu_0^\prime =1+\mu_1+\mu_2$ and
$\gamma=\frac{\lambda+\mu_2}{1+\mu_1+\mu_2}$, i.e.,
$\bar{\gamma}=1-\gamma=\frac{
\bar{\lambda}+\mu_1}{1+\mu_1+\mu_2}$. Thus, the conditions in
(\ref{KKT_1})-(\ref{KKT_2}) can be written as
\begin{align}
\hspace{-0.5cm}\mu^\prime (\bbk_1^*+\bbk_2^*+\bbsigma_1)^{-1}+\bbm_1&=\mu_0^\prime\gamma(\bbk_1^*+\bbk_2^*+\bbsigma_1)^{-1} \nonumber\\
&\hspace{-2cm}+\mu_0^\prime \bar{\gamma}(\bbk_1^*+\bbk_2^*+\bbsigma_2)^{-1}+\bbm_S \label{KKT_again_1_1}\\
\mu^\prime(\bbk_2^*+\bbsigma_2)^{-1}+\bbm_2&=\mu^\prime
(\bbk_2^*+\bbsigma_1)^{-1}+\bbm_1 \label{KKT_again_2_1}
\end{align}
which are the necessary conditions that the following problem
needs to satisfy
\begin{align}
\max_{(R_0,R_1,R_2)\in \mathcal{R}_{12}^{\rm NS-DPC}(\bbs)}
\mu_0^\prime R_0+ \mu^{\prime} (R_1 +R_2) \label{intuition}
\end{align}
On the other hand, due to
(\ref{hannan_did_it_1})-(\ref{hannan_did_it_2}), we know that the
solution of (\ref{intuition}) gives us the private message sum
rate capacity for a given common message rate, i.e., the points
that achieve the maximum in (\ref{intuition}) are on the boundary
of the capacity region $\mathcal{C}^{\rm NS}(\bbs)$. Furthermore,
the maximum value in (\ref{intuition}) can also be achieved by
using the other possible encoding order, i.e.,
\begin{align}
&\max_{(R_0,R_1,R_2)\in \mathcal{R}_{12}^{\rm NS-DPC}(\bbs)}
\mu_0^\prime R_0+ \mu^{\prime} (R_1 +R_2)\nonumber\\
&= \max_{(R_0,R_1,R_2)\in \mathcal{R}_{21}^{\rm NS-DPC}(\bbs)}
\mu_0^\prime R_0+ \mu^{\prime} (R_1 +R_2)
\end{align}
Thus, this discussion reveals that there is a one-to-one
correspondence between any rate triple on the boundary of
$\mathcal{R}_{12}^{\rm S-DPC}(\bbs)$ and the private messages sum
rate capacity points on $\mathcal{C}^{\rm NS}(\bbs)$. Hence, the
boundary of $\mathcal{R}_{12}^{\rm S-DPC}(\bbs)$, similarly
$\mathcal{R}_{21}^{\rm S-DPC}(\bbs)$, can be constructed by
considering the private messages sum rate capacity points on
$\mathcal{C}^{\rm NS}(\bbs)$. This connection between the private
messages sum rate capacity points and the boundaries of
$\mathcal{R}_{12}^{\rm S-DPC}(\bbs)$, $\mathcal{R}_{21}^{\rm
S-DPC}(\bbs)$ intuitively explains the two facts suggested by
Theorem~\ref{theorem_main_result}: i) The achievable rate region
for the Gaussian MIMO broadcast channel with common and
confidential messages is invariant with respect to the encoding
order, i.e., $\mathcal{R}_{12}^{\rm
S-DPC}(\bbs)=\mathcal{R}_{21}^{\rm S-DPC}(\bbs)$ because the
boundaries of these two regions correspond to those points on the
DPC region for the Gaussian MIMO broadcast channel with common and
private messages, for which encoding order does not matter either.
ii) We can obtain the entire capacity region of the Gaussian MIMO
broadcast channel with common and confidential messages, although
the capacity region of its non-confidential counterpart is not
known completely. The reason is that the boundary of the capacity
region of the Gaussian MIMO broadcast channel with common and
confidential messages comes from those points on the boundary of
the DPC region of its non-confidential counterpart, which are
known to be tight, i.e., on the boundary of the capacity region of
the Gaussian MIMO broadcast channel with common and private
messages.

\section{Conclusions}
We study the Gaussian MIMO broadcast channel with common and
confidential messages, and obtain the entire capacity region. We
show that a variant of the S-DPC scheme proposed
in~\cite{Ruoheng_MIMO_BC} is capacity-achieving. We provide the
converse proof by using channel enhancement~\cite{Shamai_MIMO} and
an extremal inequality from~\cite{Liu_Compound}. We also
investigate the connections between the Gaussian MIMO broadcast
channel with common and confidential messages and its
non-confidential counterpart
to provide further insight into capacity result we
obtained.

\bibliographystyle{unsrt}
\bibliography{IEEEabrv,references2}
\end{document}